\documentclass[aps,pra,nofootinbib,
twocolumn,showpacs,superscriptaddress]{revtex4}  
\usepackage{graphicx,dcolumn,bm,amsmath,amssymb} 
\usepackage{epstopdf}
\usepackage{wrapfig}
\usepackage{longtable}
\DeclareGraphicsExtensions{.eps}
\usepackage{color}
\begin{document}

\title{Optimized multibeam configuration for observation of QED cascades}
\author{E.G. Gelfer}\affiliation{National Research Nuclear University ``MEPhI'' (Moscow Engineering Physics Institute), 115409 Moscow, Russia}
\author{A.A. Mironov}\affiliation{National Research Nuclear University ``MEPhI'' (Moscow Engineering Physics Institute), 115409 Moscow, Russia}
\author{A.M. Fedotov}\affiliation{National Research Nuclear University ``MEPhI'' (Moscow Engineering Physics Institute), 115409 Moscow, Russia}
\author{V.F. Bashmakov}\affiliation{Institute of Applied Physics, Russian Academy of Sciences, 603950 Nizhny Novgorod, Russia}\affiliation{ Lobachevsky National Research University of Nizhni Novgorod, 603950 Nizhny Novgorod, Russia}
\author{E.N. Nerush}\affiliation{Institute of Applied Physics, Russian Academy of Sciences, 603950 Nizhny Novgorod, Russia}\affiliation{ Lobachevsky National Research University of Nizhni Novgorod, 603950 Nizhny Novgorod, Russia}
\author{I.Yu. Kostyukov}\affiliation{Institute of Applied Physics, Russian Academy of Sciences, 603950 Nizhny Novgorod, Russia}\affiliation{ Lobachevsky National Research University of Nizhni Novgorod, 603950 Nizhny Novgorod, Russia}
\author{N.B. Narozhny}\affiliation{National Research Nuclear University ``MEPhI'' (Moscow Engineering Physics Institute), 115409 Moscow, Russia}
\begin{abstract}
QED cascades in intense electromagnetic field can occur if the dynamical quantum parameter $\chi$ of a seed electron, which in Compton units coincides with the electron proper acceleration, attains the order of unity. We derive general expression for $\chi$ of an initially slow electron in an arbitrary electromagnetic field for a time range $t\ll 1/\omega$, where $\omega$ is the field carrier frequency. Using this formula, we consider a special field configuration of multiple colliding focused laser beams and optimize it to provide cascade development at laser power below $10$~PW and intensity of the order of $10^{23}$W/cm$^2$. Such parameters of the beams will be obtained with a new generation of laser facilities, particularly the ELI Beamlines, in the coming years.
\end{abstract}
\maketitle

\section{Introduction}
During the last two decades one could observe a remarkable progress in laser technology. Since the invention of the CPA technique \cite{mourou} in 1985, the maximum  laser intensity has been increased by seven orders of magnitude reaching nowadays the value $2\cdot 10^{22}$W/cm$^2$ \cite{yanovsky}. Moreover, the announced new generation of $10$~PW laser facilities like ELI Beamlines, which is now under construction in Prague, Czech Republic \cite{ELI}, or VULCAN in UK \cite{vulcan}, will provide intensity above $10^{23}$W/cm$^2$. Therefore physics in the presence of such ultra-strong fields is now coming upon a focus of attention. One of the most striking effects in this area is electron-positron pair creation from vacuum \cite{pairs1,nikishov,pairs2,fedotovlp,narprl}. However, according to recent calculations \cite{narprl}, spontaneous pair creation from vacuum would require an optical laser of the power above $500$~PW, still far from the experimental possibilities of the nearest future. 

Another important example of ultra-strong field physics phenomena are quantum electrodynamical (QED) cascades \cite{kirk,limitations,elkina,nerush,duclous,polarization}. In the present paper we consider cascades seeded by electrons, that are initially at rest at the focus of intense laser field \footnote{Note that it is also possible to initiate QED cascades in a laser field by seed hard photons or high-energy electron beam \cite{mironov}, the latter similar to the SLAC experiment \cite{slac}.}. They are accelerated by laser field emitting hard photons, which in turn create further electron-positron pairs. The created electrons and positrons are again accelerated by the field thus emitting hard photons, which again create new pairs and so on. Such an avalanche-like process finishes by escape of the particles from the focal region. 

Observation of QED cascade production in a laboratory is important for both fundamental physics and testing the QED principles. However, numerical simulations performed in the previous works \cite{kirk,elkina} for some particular models of laser pulses have indicated that the threshold intensity required for cascade development would be of the order or above $10^{24}$W/cm$^2$. In the latter case this may still exceed the capabilities of such facilities as e.g. the ELI Beamlines. Our goal is to optimize the experimental setup, in particular the field distribution at the focus, thus suggesting configuration that decreases the threshold intensity required for QED cascade production. 

Consider the QED field strength parameter $\chi$ \cite{prob}
\begin{equation}\label{chi}
\chi=\frac{e\hbar}{m^3 c^5}\varepsilon F_\bot,
\end{equation}
where $m$ and $e$ are the electron mass and charge, $\varepsilon$ is the particle energy and $\mathbf{F}_\bot$ is  the component of the Lorentz force per unit charge $\mathbf{F}=\mathbf{E}+[\mathbf{v}\times \mathbf{B}]$ transverse with respect to the particle velocity $\mathbf{v}$. In the framework of a locally constant field approximation \cite{pairs2,narprl,limitations} the parameter (\ref{chi}) evolves in between the hard photon emission and pair creation events according to the classical equations of motion and its instant value defines the probabilities of these processes \cite{prob}
\begin{equation}\nonumber
W_{e,\gamma}(\chi_{\gamma,e}\gg1)\sim\frac{\alpha m^2 c^4}{\hbar\varepsilon_{e,\gamma}}\chi_{\gamma,e}^{2/3}, \quad W_e(\chi_\gamma\lesssim1)\sim e^{-\frac{8}{3\chi_\gamma}},
\end{equation}
where $W_e$ and $W_\gamma$ are the probabilities of pair creation and hard photon emission, respectively, $\varepsilon_{e,\gamma}$ is the energy of radiating electron or pair creating photon and $\alpha=e^2/\hbar c$ is the fine structure constant. Since $\chi_\gamma$ is less than $\chi_e$ of a radiating electron and pair creation is exponentially suppressed at $\chi_\gamma\lesssim 1$, we have to find such a field configuration that provides the most rapid growth of $\chi$ \footnote{Herein and onwards we abbreviate $\chi_e$ back to $\chi$.}. Then we calculate numerically the threshold  intensity for QED cascade development in such an optimal field configuration. 

\section{Growth of parameter $\chi$}

Assume that a seeding particle is placed initially at rest at the focus of the laser field. In order to optimize field configuration for cascade production, we derive first a formula for the time dependence $\chi(t)$ near the focus, i.e. in a time range $t\ll 1/\omega$, where $\omega$ is the field carrier frequency. In this case the electric and magnetic fields can be expanded into Taylor series
\begin{equation}\label{EB}\nonumber
\mathbf{E}(x)\approx\mathbf{E}_0+\mathbf{E}_{,k}(0)x^k,\quad \mathbf{B}(x)\approx\mathbf{B}_{,k}(0)x^k
\end{equation}
with respect to time and coordinates $x^k=\{ct,\vec{x}\}$. Since a standing wave composed of two identical counterpropagating laser beams produces cascades more efficiently than a single beam of the same total power \cite{kirk}, we assume in advance that the electric field at the center of the focal region reaches maximum and the magnetic field vanishes. Besides, in what follows we assume that the laser field is so strong that particles become ultrarelativistic soon after start of acceleration, $eEt_{acc}/mc\gg1$, where  $t_{acc}$ is the acceleration time, but still remains much weaker than the QED critical field, $E\ll E_{cr}=m^2c^3/e\hbar=1.3\cdot10^{16}$V/cm.

Let us find the particle momentum satisfying the equation of motion $\dot{\mathbf{p}}=e\mathbf{F}$ in the form $\mathbf{p}=\mathbf{p}^{(0)}+\mathbf{p}^{(1)}+\ldots$, where $\mathbf{p}^{(k)}=\mathcal{O}((\omega t)^k)$.
For $t\gg m/eE_0$ \footnote{We use the natural units $\hbar=c=1$ onwards.} the solution of equations of motion, accurate to the zeroth order in $\omega t$, reads
\begin{equation}\nonumber
\mathbf{p}^{(0)}=e\mathbf{E}_0 t,\quad \varepsilon^{(0)}=eE_0t,\quad \mathbf{r}^{(0)}=\mathbf{v}^{(0)}t=\frac{\mathbf{E}_0}{E_0}t.
\end{equation}

Using expressions for the fields within the first order in $\omega t$, $\mathbf{E}(x)=\mathbf{E}_0+\mathbf{E}'t$, $\mathbf{B}(x)=\mathbf{B}'t$,
where
\begin{equation}\label{deb}
\mathbf{E}'=\left\{\frac{\partial}{\partial t}+\mathbf{v}^{(0)}\nabla\right\} \mathbf{E}(0), \quad \mathbf{B}'=\left\{\frac{\partial}{\partial t}+\mathbf{v}^{(0)}\nabla\right\} \mathbf{B}(0),
\end{equation}
we can solve the first order equation of motion $d\mathbf{p}^{(1)}/dt=e\mathbf{F}^{(1)}=e{\mathbf{f}^{(1)}}t$, where
\begin{equation}\label{f}
{\mathbf{f}^{(1)}}=\mathbf{E}'+[\mathbf{v}^{(0)}\times\mathbf{B}'],
\end{equation}
and obtain for the particle velocity $\mathbf{v}=\mathbf{p}/\varepsilon$,
\begin{equation}
\begin{split}
\mathbf{v}\approx\mathbf{v}^{(0)}+\frac{\left\{\mathbf{f}^{(1)}-\mathbf{v}^{(0)}(\mathbf{v}^{(0)}\mathbf{f}^{(1)})\right\}}{2E_0}t.
\end{split}
\end{equation}

Since the magnitude of the transverse component of the Lorenz force is equal to $F_\bot=|[\mathbf{F\times v}]|=ef_\bot^{(1)} t/2$, where $f_\bot^{(1)}$ is the component of (\ref{f}) transverse to the initial electric field $\mathbf{E}_0$, we arrive at the formula
\begin{equation}\label{chipar}
\chi(t)=\frac{e^2E_0f_\bot^{(1)}}{2m^3}t^2.
\end{equation}
The result (\ref{chipar}) is in agreement with  special cases of rotating purely electric field and a standing plane wave obtained previously in Refs.~\cite{limitations} and \cite{polarization}. 

Hence, in order to optimize field configuration for cascade production we must increase the coefficient at $t^2$ in the formula (\ref{chipar}) as much as possible. 

\section{Colliding beams}

\begin{figure}[h!]
\begin{center}
\includegraphics[width=0.5\columnwidth]{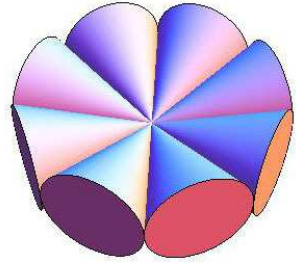}
\end{center}
\caption{\label{8beams}
(Color online). Allocation of $n=8$ tightly focused beams colliding in a plane.}
\end{figure}

A method for reducing the total threshold power for spontaneous pair creation from vacuum via multibeam technology was suggested in Ref.~\cite{narprl}, where it was clearly demonstrated how the key parameter of that problem, the electric field strength in the focus, can be increased  considerably by splitting a laser beam into several pairs of linearly polarized colliding beams of the same total power. However, in our case of cascade production linear polarization is expected to be unfavorable \cite{polarization,bulanov}. Therefore we modify that scheme not restricting polarization of individual beams a priory.

Let us consider an even number $n\ge 2$ of focused beams propagating in $xy$ plane at angles
\begin{equation}\nonumber
\varphi_j=2\pi(j-1)/n,\quad j=1\ldots n
\end{equation} 
with the $x$ axis (see Fig.~\ref{8beams}) and optimize their polarization for cascade production. For not too strong focused beams their field near the focus can be well approximated by plane waves
\begin{equation}\label{aplane}
\mathbf{A}_j=-i{\cal A}_0\boldsymbol{\epsilon}^je^{-i\omega(t-\mathbf{r}\mathbf{\hat{n}}^j)}.
\end{equation}
Here $\boldsymbol{\epsilon}^j=\boldsymbol{\epsilon}^j_1-i\boldsymbol{\epsilon}^j_2$, $\boldsymbol{\epsilon}^j_1\bot\boldsymbol{\epsilon}^j_2$ are the individual polarization vectors, which we do not require in advance to be normalized, and $\mathbf{\hat{n}}^j\bot\boldsymbol{\epsilon}^j$ are the unit vectors directed along the focal axis of the beams. In case all the beams are focused weakly, the power of each of the beams can be estimated as $P_j={\cal A}_0^2|\boldsymbol{\epsilon}^j|^2/\varkappa^2$, where $\varkappa=\sqrt{8\pi/\omega^2S_\perp}$ is a geometric factor depending on the beam shape, $S_\perp\gg\omega^{-2}$ is the focal cross section area of the beam. The same kind of formula remains true for a tight focusing case as well, but the geometric factor in such a case is modified (see below). Below, we prefer to fix the total power $P=\sum_{j=1}^n P_j$ of the beams, in terms of which
\begin{equation}\label{A0}
{\cal A}_0=\varkappa\frac{\sqrt{P}}{\epsilon},\quad \epsilon^2=\sum\limits_j \left[(\epsilon^j_1)^2+(\epsilon^j_2)^2\right].
\end{equation}

Due to transversality of the plane waves (\ref{aplane}), all the terms containing space derivatives in (\ref{deb}) vanish within this approximation, so that $\mathbf{B}'=-i\omega\mathbf{B}(0)=0$. Substituting $\mathbf{E}=-\sum\limits_j d\mathbf{A}_j/dt$ into (\ref{chipar}) and taking into account (\ref{A0}), for $n$ colliding beams we obtain
\begin{equation}
\label{chin}
\chi(t)=\frac{e^2\omega^3P{\cal F}_n\varkappa^2}{2m^3}t^2,
\end{equation}
where the factor
\begin{equation}\label{F}
{\cal F}_n=\frac{\epsilon_1\epsilon_2}{\sum\limits_{j=1}^n\left[(\epsilon_1^j)^2+(\epsilon_2^j)^2\right]},\quad \boldsymbol{\epsilon}_{1,2}=\sum\limits_{j=1}^n\boldsymbol{\epsilon}^j_{1,2}
\end{equation}
totally determines dependence of $\chi$ on the number of beams and their polarization. 

Evidently, in order to maximize the expression (\ref{F}) all the vectors $\boldsymbol{\epsilon}^j_1$ should be parallel each other. Since the $xy$-plane had been chosen as the plane of beam propagation, we assume thus that $\boldsymbol{\epsilon}^j_1$ are all directed along the $z$-axis, $\boldsymbol{\epsilon}^j_1=g_j \mathbf{\hat{z}}$, where $\mathbf{\hat{z}}$ is the corresponding versor. Also, since the electric field in the beam is transverse to the wave propagation direction $\mathbf{\hat{n}}^j$\footnote{Note that since the magnetic field in the focus is assumed to be equal to zero, we must impose a constraint $\boldsymbol{\epsilon}^j=\boldsymbol{\epsilon}^{n/2+j}$, or just $h_j=-h_{n/2+j}$, for all $j=1\ldots n/2$.}, we can assume that $\boldsymbol{\epsilon}^j_2=-h_j\sin\varphi_j \mathbf{\hat{x}}+h_j\cos\varphi_j \mathbf{\hat{y}}$.

Due to the symmetry of the problem the polarization vector of the total electric field can lie in any plane parallel to $z$-axis. Let it be the $yz$- plane. Then 
\begin{equation}\label{Ff}
{\cal F}_n=\frac{\left(\sum\limits_{j=1}^n h_j\cos\varphi_j\right)\left( \sum\limits_{j=1}^n g_j\right)}{\sum\limits_{j=1}^n(h_j^2+g_j^2)}=\frac{(\boldsymbol{\alpha H})(\boldsymbol{\beta G})}{\lVert\boldsymbol{H}\rVert^2+\lVert\boldsymbol{G}\rVert^2},
\end{equation}
where, to clarify further manipulations, in the last representation we have introduced four formal $n$-vectors $\boldsymbol{G}=\{g_j\}$, $\boldsymbol{H}=\{h_j\}$, $\boldsymbol{\alpha}=\{\cos\varphi_j\}$ and $\boldsymbol{\beta}$, the latter with all $n$ components equal to $1$.  

Now it follows from the chain of inequalities\footnote{Here we apply by turn the inequality of arithmetic and geometric means and then the Cauchy-Bunyakovsky inequality.}
\begin{equation}\label{Fmax}
{\cal F}_n\le\frac{1}{2}\frac{(\boldsymbol{\beta G})}{\lVert\boldsymbol{G}\rVert}\frac{(\boldsymbol{\alpha H})}{\lVert\boldsymbol{H}\rVert}\le\frac{\lVert\boldsymbol{\alpha}\rVert \lVert\boldsymbol{\beta}\rVert}{2}=\frac{n}{2\sqrt{2}},
\end{equation}
that ${\cal F}_n$ reaches its maximum value when $\mathbf{G}\parallel\boldsymbol{\beta}$, $\mathbf{H}\parallel\boldsymbol{\alpha}$ and $\lVert\mathbf{G}\rVert=\lVert\mathbf{H}\rVert$. For $n>2$ these conditions are equivalent to
\begin{eqnarray}\label{ez}
g_j=g_1,\quad \boldsymbol{\epsilon}_1=ng_1\mathbf{\hat{z}},\quad h_j=h_1\cos\varphi_j,\\
\label{exy}
\boldsymbol{\epsilon}_2=h_1\mathbf{\hat{y}}\sum\limits_{j=1}^n\cos^2\varphi_j=\frac{n}{2}h_1\mathbf{\hat{y}}.
\end{eqnarray} 
In particular,
\begin{equation}\label{21}
\epsilon_1:\epsilon_2=\sqrt{2}:1,
\end{equation}
i.e. the optimal polarization of the total field is elliptical with the semi-major to semi-minor axis ratio equal to $\sqrt{2}:1$. The polarization vectors of the individual beams in such an optimal case are 
\begin{equation}\label{optpol}
\boldsymbol{\epsilon}_j=\{i\sqrt{2}\sin\varphi_j\cos\varphi_j,-i\sqrt{2}\cos^2\varphi_j,1\}.
\end{equation}
In case $n=2$ the factor $2$ does not appear in denominator of Eq.~(\ref{exy}), hence the optimal polarization is circular, $\epsilon_1:\epsilon_2=1:1$.

\section{Numerical results}

\renewcommand{\arraystretch}{1.8}
\renewcommand{\tabcolsep}{0.4cm}
\begin{table}[h!]
\caption{\label{Tab1}Power and intensity thresholds for generation of QED cascades in configurations with $n$ colliding beams ($\omega=1eV$, $\xi=8$).}
\label{tabular:timesandtenses}
\begin{center}
\begin{tabular}{|c|c|c|c|c|}
\hline
$n$ & 2 & 4 & 6 & 8 \\ \hline
$P_{th}, \text{PW}$ & 23 & 14 & 10 & 7.9 \\ \hline
$I_{th}, 10^{23}\text{W/cm}^2$ & 17 &10 &7.3 &5.6 \\
\hline
\end{tabular}
\end{center}
\end{table}
\begin{figure}[h!]
\begin{center}
\includegraphics[width=0.9\columnwidth]{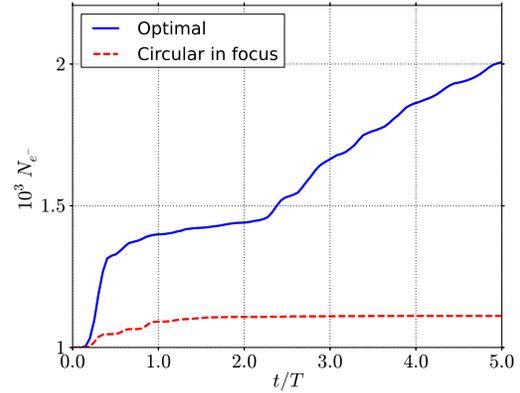}
\end{center}
\caption{\label{Nt}
(Color online) Number of electrons in a cascade versus time for optimal (solid) and circular (dashed) polarization of the total field ($n= 8$, $\xi=8$, $\omega=1$eV, $P=7.9\cdot10^{15}$W, $T$ is the laser period).}
\end{figure}

\begin{figure*}[t]
\begin{minipage}[h]{0.32\linewidth}
\center{\includegraphics[width=\linewidth]{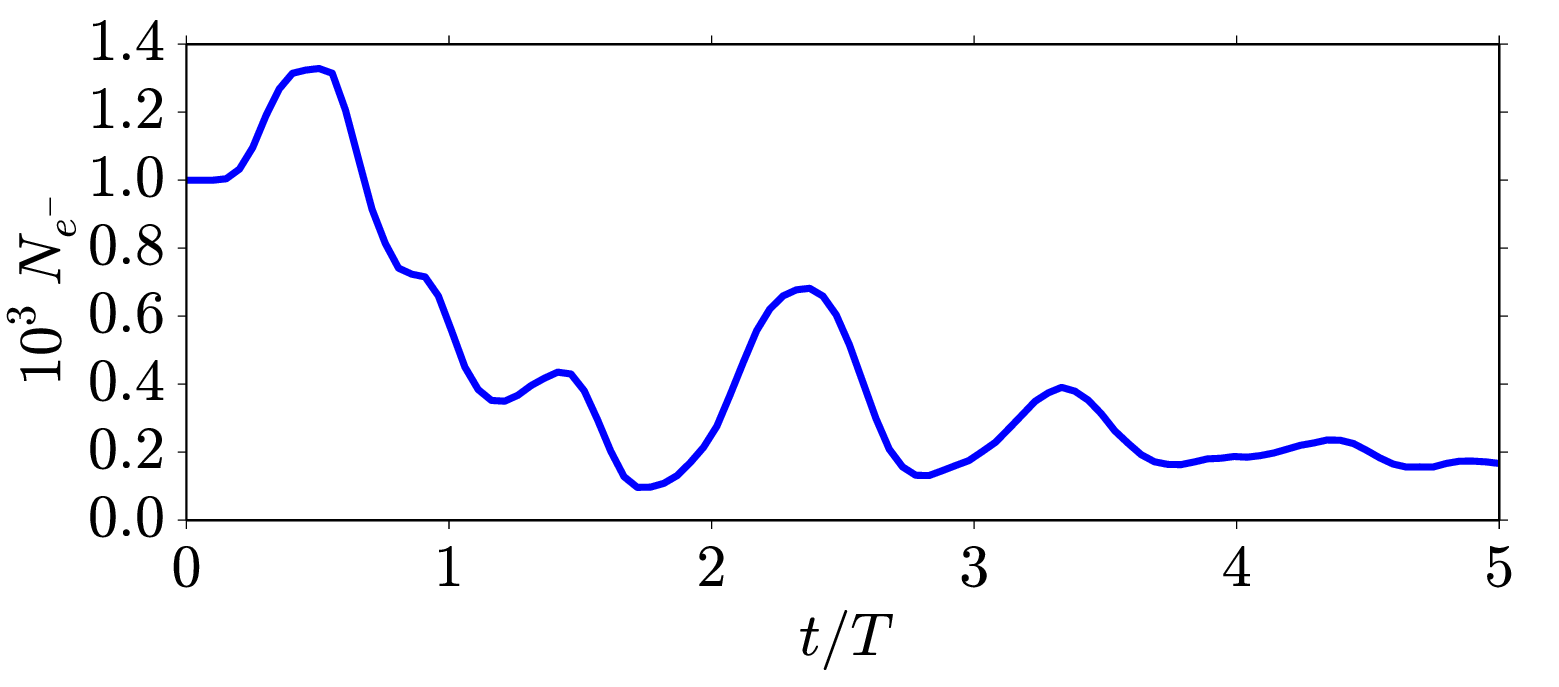} \\ a)}
\end{minipage}
\hfill
\begin{minipage}[h]{0.32\linewidth}
\center{\includegraphics[width=\linewidth]{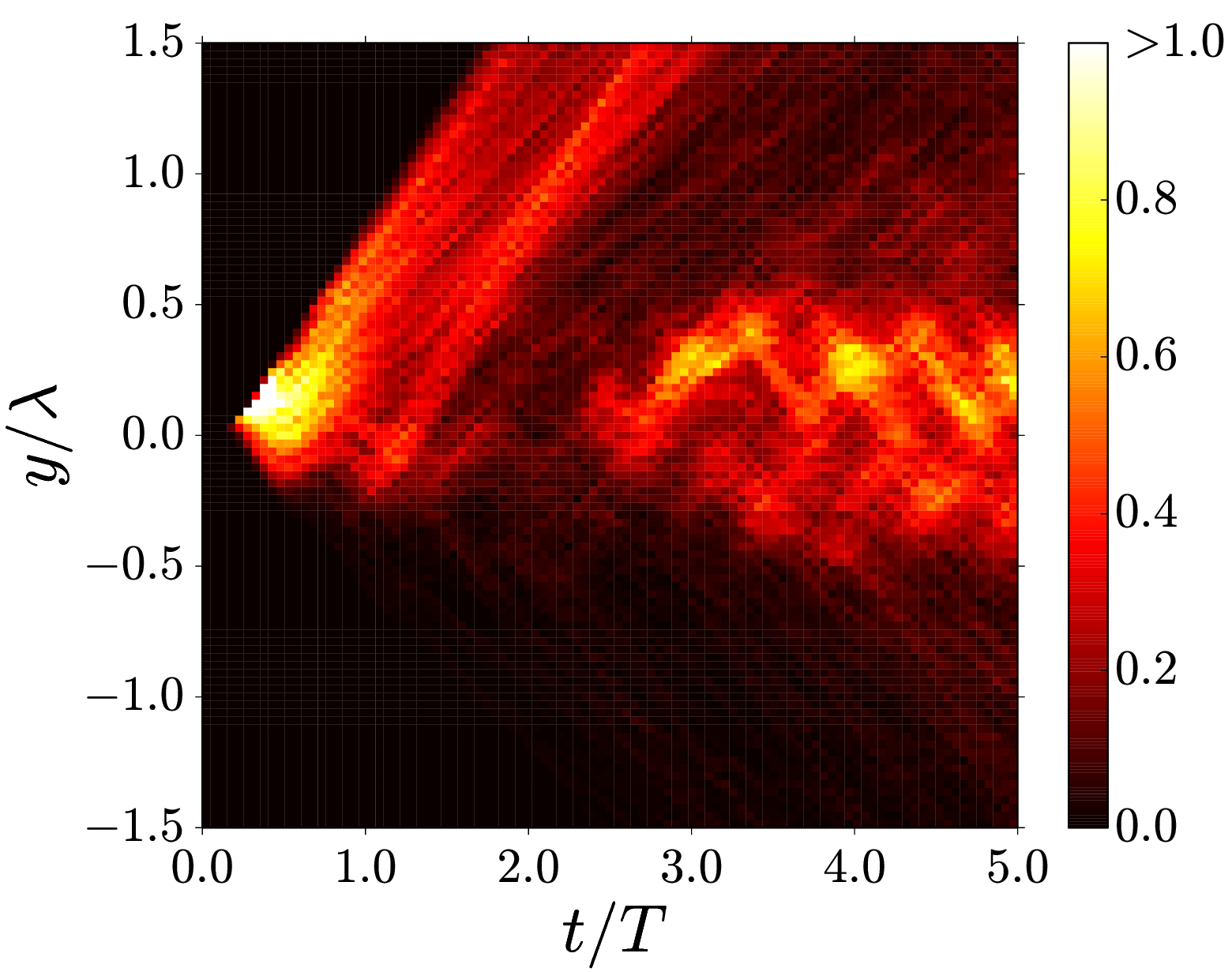} \\ b)}
\end{minipage}
\hfill
\begin{minipage}[h]{0.32\linewidth}
\center{\includegraphics[width=\linewidth]{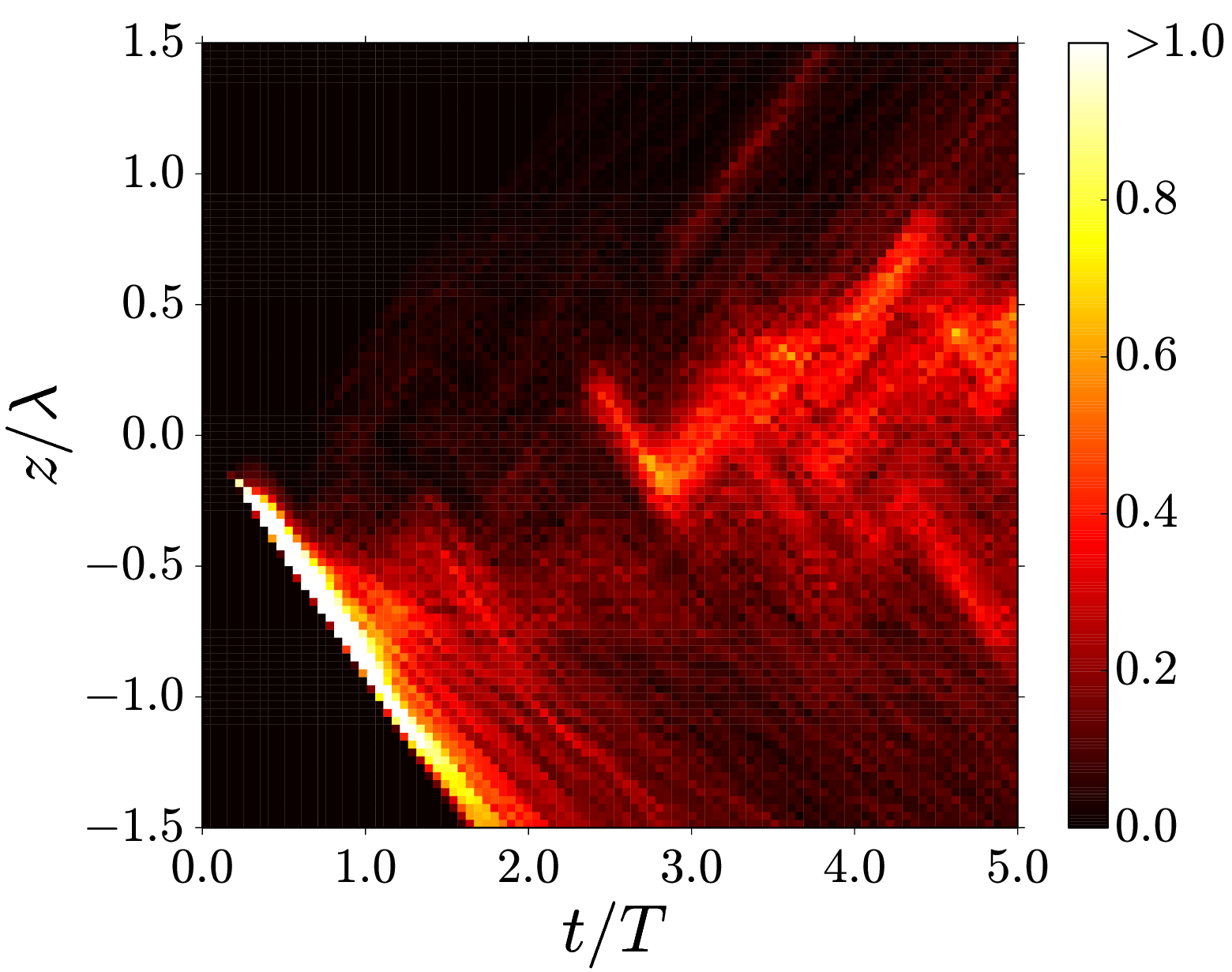} \\ c)}
\end{minipage}
\caption{(Color online). (a) number of electrons at a distance smaller than $\lambda$ from the origin; (b), (c) -- dimensionless linear densities $\frac{1}{N_0} \frac{dN}{dy} \lambda$ of created positrons along $y$- and $z$-axes. The parameters are the same as in Fig.~\ref{Nt}.}
\label{positrons}
\end{figure*}

\begin{figure}[h]
\begin{center}
\includegraphics[width=0.9\columnwidth]{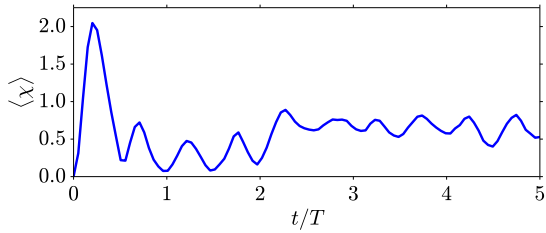}
\end{center}
\caption{\label{chif}
(Color online). The mean value of $\chi$ for electrons located at a distance smaller than $\lambda$ from the origin. The parameters are the same as in Fig.~\ref{Nt}.}
\end{figure}

The threshold power for cascade generation can be found by numerical simulations. In fact, QED cascades do not possess a natural sharp threshold, so that we define it formally as creation of just a single pair on average by each seed electron during passage of laser pulses through the focus. As a more realistic model we use a model of the focused laser beam suggested in Refs.~\cite{narfof,fedotovlp}
\begin{equation}\label{afed}
\begin{split}
\mathbf{A}_j=&\frac{4\xi^2\sqrt{P}e^{-i\omega(t-r^j_\parallel)}}{\epsilon\omega\sqrt{2\xi-3}(\xi+i\omega r^j_\parallel)^2}\exp\left[-\frac{(\omega r^j_\bot)^2}{2(\xi+i\omega r^j_\parallel)}\right]\\
&\times\left(\boldsymbol{\epsilon}^j-\frac{\omega^2{([\boldsymbol{\epsilon}^j\times\hat{\mathbf{n}}^j]\cdot \mathbf{r}^j_\bot)[\hat{\mathbf{n}}^j\times \mathbf{r}^j_\bot]}}{\xi+i\omega r^j_\parallel}\right)
\end{split}
\end{equation}
which is an approximate solution of Maxwell equations near the focal region if $\xi\gg1$. Here, $r^j_\parallel$ and $\mathbf{r}^j_\bot$ are the  components of the radius vector, parallel and perpendicular to $\hat{\mathbf{n}}^j$. The parameter $\xi=\omega L/2$, where $L$ is the diffraction length, controls the degree of beam focusing; $\mathbf{\hat{n}}^j$, $\boldsymbol{\epsilon}^j$ and $\epsilon$ are all defined after Eq.~(\ref{aplane}); $P$ is the total power of all colliding beams.  

It follows from (\ref{chin}) that for given total power, the geometric factor $\varkappa$ (i.e., focusing) should be maximized. Note that in modern laser facilities the beam can be focused so tightly, that a half of the beam power is passing through a circle of the wavelength diameter in the focal plane \cite{yanovsky}. Computing numerically the power of a beam over a circle of the diameter $\lambda=2\pi/\omega$ around the focus, we obtain, that the condition
\begin{equation}\nonumber
\frac{1}{8\pi}\mathrm{Re}\int\limits_0^{2\pi}d\phi\int\limits_0^{\lambda/2}\rho d\rho\,\left.\mathbf{\hat{n}}^j\cdot[\mathbf{E}_j\times \mathbf{B}_j^*]\right|_{r^j_\parallel=0}=\frac{P_j}{2},
\end{equation}
where $P_j$ is the power of the $j$-th beam, is satisfied if the focusing parameter $\xi\simeq 8$. Here $\rho$ and $\phi$ are the polar coordinates in the plane $r^j_\parallel=0$. 

It is also worth taking into account that there exist bounds for a number of beams that can be allocated concurrently in a plane without overlapping \cite{narprl,Gonoskov}. At far distance from the focal region the field of each of the beams is almost confined to a cone with aperture $\theta$, which is simply related to our focusing parameter $\xi$. Indeed, for $|\mathbf{r}|\gg \xi/\omega$ the energy of the field is proportional to $e^{-\xi r_\bot^2/r_\parallel^2}$. Therefore we can estimate the aperture by $\theta\sim 2\arctan1/\sqrt{\xi}$. For $\xi=8$, $\theta=0.216\pi>\pi/5$, hence the number of beams that can be allocated without overlapping in the $xy$-plane must be less than $10$. Hence, for $\xi=8$ the maximal even number of beams is $8$ as in Fig.~\ref{8beams}.

For numerical simulations we have used two independent codes. The first one \cite{mironov} is a realization of Monte-Carlo (MC) algorythm, in which we assume that electrons and positrons are moving classically between the emission events, which are implemented by the corresponding event generators similar to Refs.~\cite{elkina,duclous}. The second one is a more general 3D Particle-in-Cell-Monte-Carlo (PIC-MC) code for simulation of laser-plasma interactions with quantum radiation effects \cite{nerush,polarization,pic}. Actually, in our problem the electron-positron plasma remains dilute and the results obtained with both codes are in good agreement. In this paper we present the results obtained with the first (pure MC) code.

Since our field model (\ref{afed}) corresponds to a field unbounded in time, we take into account finiteness of pulse duration artificially by limiting the simulation time. The simulation time was chosen to be $5$ laser periods $T$ as is typical for modern laser facilities \cite{yanovsky}. In simulations, the seed particles were represented by a bunch of $N_0=10^3$ electrons placed initially at rest near the center $\mathbf{r}=0$ of the focus. The laser carrier wavelength was taken to be $\lambda=1.24 \mu$m, which corresponded to laser photon energy $1$eV and the focusing parameter was chosen as $\xi=8$ for the reasons discussed above. 

The resulting power and intensity thresholds  obtained by numerical simulation of QED cascade development in the cases of $n=2$, $4$, $6$, and $8$ beams with optimal polarization (\ref{optpol}), colliding in a plane are presented in Table~\ref{Tab1}. One can observe that both the laser power and intensity required for cascade production decrease significantly due to a multibeam setup. Cascade dynamics for the cases of optimal elliptic polarization and circular polarization of the total field for the case of $n=8$ beams is displayed on Fig.~\ref{Nt}. The figure clearly demonstrates the advantage of the suggested optimal polarization.

Besides, it is clear from Fig.~\ref{Nt} that in the case of optimal polarization cascade development is splitted into two stages, which differ by the slope of the curve. At the first stage the seeding particles are accelerated by the field, create pairs and escape the focal region. Almost all created particles also leave the focal region with a speed close to the speed of light, see Fig.~\ref{positrons}. Still, a small fraction of slow particles remains in the focal region being accelerated by the field. At time $t\simeq 2.25 T$ the mean value of the parameter $\chi$ for particles at the focus becomes large enough to initiate the second stage of cascade development, see Fig.~\ref{chif}.

\section{Conclusion}
We have demonstrated that QED cascades can arise in laser-target interaction for laser power below $10$~PW and intensity below $10^{24}$W/cm$^2$. This is realizable with a multibeam setup, i.e. splitting of the original laser pulses into several pairs of colliding beam. In order to determine how to arrange the beams, we have derived an analytical formula for growth of the parameter $\chi(t)$, which determines possibility of QED cascade generation. This allowed us to identify the optimal beam polarization with the most rapid growth of $\chi(t)$ at the focus. For such an optimal configuration of $n$ beams and fixed total laser power, $\chi$ is proportional to $n$. However, for our extreme choice of the focusing parameter, which corresponds to the state-of-the-art experimental capabilities, the maximal number of beams that can be brought together in a planar configuration of Fig.~\ref{8beams} is $n=8$. Thus we have next computed the QED cascade thresholds for the particular case of $n=8$ optimally polarized beams. According to our simulations, in this case the threshold total power $P_{th}=7.9$~PW and the threshold intensity $I_{th}=5.6\cdot10^{23}$W/cm$^2$. Such parameters had been previously announced as being achievable with the new generation of laser facilities \cite{ELI,vulcan} in the nearest future.

\section{Acknowledgements}
\begin{acknowledgments}
This research was supported by the Government of the Russian Federation (Project No. 14.B25.31.0008), the Russian Fund for Basic Research (grants 13-02-00372 and 13-02-00886) and the RF President program for support of young Russian scientists and leading research schools (grant NSh-4829.2014.2). AAM acknowledges partial support from the Dynasty Foundation.
\end{acknowledgments}

\end{document}